\documentclass[draft]{ifacconf}

\usepackage{graphicx}      % include this line if your document contains figures
\usepackage{natbib}        % required for bibliography
\usepackage{amsmath,amssymb,amsfonts}
\usepackage[shortlabels]{enumitem}
\newtheorem{theorem}{Theorem}
\newtheorem{assumption}{Assumption}
\newtheorem{definition}{Definition}
\newtheorem{lemma}{Lemma}
\newtheorem{proposition}{Proposition}
\newtheorem{remark}{Remark}
\allowdisplaybreaks[4]

\usepackage{pgfplots}

\definecolor{tuklblue}{RGB}{0,95,140}
\definecolor{tuklred}{RGB}{185,40,25}
\pgfplotsset{compat=1.16}
\pgfplotsset{colormap={tuklcolor}{color(0cm)=(black!60!tuklblue); color(0.2cm)=(tuklblue); color(4cm)=(tuklblue!20)}}
\usepackage{subcaption}
\usetikzlibrary{cd}
\pgfplotsset{/tikz/commutative diagrams/math mode=false}

\newcommand{\N}{\ensuremath{\mathbb{N}}}	% natural numbers
\newcommand{\R}{\ensuremath{\mathbb{R}}}	% real numbers
\newcommand{\Cont}{\ensuremath{\mathcal{C}}}	% space of continuous functions
\newcommand{\ball}{\ensuremath{\mathcal{B}}}	% open ball
\newcommand{\diff}{\ensuremath{\textup{d}}}	% differential
\newcommand{\eps}{\ensuremath{\varepsilon}}	% abbreviation for \varepsilon
\newcommand{\norm}[1]{\ensuremath{\left\|}#1\ensuremath{\right\|}}	% norm
\newcommand{\abs}[1]{\ensuremath{\left|}#1\ensuremath{\right|}}	% absolute value
\newcommand{\ph}{\, \cdot \,}				% placeholder with dot
\newcommand{\image}{\operatorname{im}}		% image
\newcommand{\lin}{\mathcal{L}}

\newcommand{\para}[1]{\left(#1\right)}		% parantheses ()
\newcommand{\braces}[1]{\left\{#1\right\}}	% braces {}

\newcommand{\PC}{\ensuremath{\mathcal{PC}}}	% space of piecewise continuous functions
\newcommand{\V}{V\!\para{t,x(t)}}
\newcommand{\Vcand}{V_{\operatorname{cand}}}

\begin{document}
\begin{frontmatter}

\title{Construction of time-varying ISS-Lyapunov Functions of Impulsive Systems\thanksref{footnoteinfo}}%\thanksref{footnoteinfo}} 

\thanks[footnoteinfo]{The work of Saeed Ahmed was supported by the Dutch
Research Council (NWO) under Grant ESI.2019.005.}

\author[First]{Patrick Bachmann} 
\author[Second]{Saeed Ahmed} 

\address[First]{Institute of Mathematics, University of W{\"u}rzburg, Germany (e-mail: patrick.bachmann@uni-wuerzburg.de).}
\address[Second]{Jan C. Willems Center for Systems
and Control, ENTEG, Faculty of Science and Engineering, University of Groningen, 9747 AG Groningen, The Netherlands (e-mail: s.ahmed@rug.nl)}

\begin{abstract}
Time-varying ISS-Lyapunov functions for impulsive systems provide a necessary and sufficient condition for ISS. This property makes them a more powerful tool for stability analysis than classical candidate ISS-Lyapunov functions providing only a sufficient ISS condition. Moreover, time-varying ISS-Lyapunov functions cover systems with simultaneous instability in continuous and discrete dynamics for which candidate ISS-Lyapunov functions remain inconclusive. The present paper links these two concepts by suggesting a method of constructing time-varying ISS-Lyapunov functions from candidate ISS-Lyapunov functions, thereby effectively combining the ease of construction of candidate ISS-Lyapunov functions with the guaranteed existence of time-varying ISS-Lyapunov functions.
\end{abstract}


\end{frontmatter}

\renewcommand{\labelenumi}{(\roman{enumi})}
% !TeX spellcheck = en_US
\section{Introduction}
Impulsive systems are a class of hybrid dynamical systems that combine continuous behavior with abrupt changes of state, often related by the \emph{flow} and the \emph{jump}, respectively. The flow is typically described by an ordinary differential equation or a partial differential equation. Impulsive systems find their application in various real-world applications; see \cite{Rivadeneira2015}, \cite{Antunes2013}, and \cite{Guan2012}. 

An essential property of dynamical systems in practice is their sensitivity to external perturbations. 
The notion of input-to-state stability (ISS), introduced by \cite{Sontag1989}, guarantees a certain tolerance to such inputs and is therefore helpful for classifying a system's behavior. A well-established tool for proving ISS is the ISS-Lyapunov function. The framework of ISS-Lyapunov functions was modified to the \emph{candidate ISS-Lyapunov function} to cover impulsive systems  \cite[cf.][]{Hespanha2008}. See also \cite{Dashkovskiy2013} for an ISS condition for impulsive systems on an infinite-dimensional Banach space via a candidate ISS-Lyapunov function.

A candidate ISS-Lyapunov function-based analysis, however, has certain restrictions. First, it only provides a \emph{sufficient condition for ISS}. Second, it offers stability conclusions for a restricted class of systems: Either the flow behavior must be stable, and the jumps may be unstable, or the jumps ought to be stable, and the flow might be unstable. These ISS results are inconclusive for impulsive systems with simultaneous instability of the continuous and discrete dynamics. ISS of this  class of impulsive systems has received little attention in the literature. It was not until recently that \cite{Dashkovskiy2021a} treated ISS of such impulsive systems using a dwell-time approach based on the higher-order derivatives of the Lyapunov function. However, the work of \cite{Dashkovskiy2021a} also gives only a sufficient condition of ISS and does not provide a converse ISS-Lyapunov theorem. Motivated by this, we recently proposed time-varying  ISS-Lyapunov functions in implication form that provide a necessary and sufficient condition for ISS of impulsive systems over infinite-dimensional Banach spaces; see \cite{Bachmann2022}. They also apply to  systems with simultaneous instability in continuous and discrete dynamics.

In this paper, we provide a construction method for time-varying Lyapunov functions from candidate Lyapunov functions incorporating the existing well-established candidate Lyapunov theory into our newly proposed concept of time-varying Lyapunov functions. This enables us to combine the advantages of both, namely the simplicity of construction of candidate Lyapunov functions \citep[cf.][]{Dashkovskiy2013} with the guaranteed existence of a time-varying Lyapunov function.

The rest of the paper unfolds as follows. In Section~\ref{sec:preliminaries}, we provide some preliminaries, including necessary definitions and the notion of time-varying ISS-Lyapunov functions for impulsive systems. In Section~3, we provide a necessary and sufficient condition for ISS. In Section 4, we provide the main result on the construction of ISS-Lyapunov functions. We summarize our findings in Section~\ref{sec:conclusion}.

% !TeX spellcheck = en_US
\section{Preliminaries and Input-to-State Stability}\label{sec:preliminaries}
We denote the set of natural numbers by $\N$, the set of nonnegative integers by $\N_0 = \N \cup \{0\}$, the set of real numbers by $\R$, the set of nonnegative real numbers by $\R^+_0$, the space of continuous functions from normed spaces $X$ to $Y$ by $\Cont(X,Y)$, and the ball of radius $r > 0$ around $0$ by $\ball_X(r)$. Let $I = [t_0,\infty) \subset \R$ and let $S = \braces{t_n}_{n \in \N}$ be a set, which contains the elements of a strictly increasing sequence of impulse times $(t_n)_{n \in \N}$ in $(t_0, \infty)$ such that $t_i \to \infty$ for $i \to \infty$. Let $\para{X, \norm{\ph}_X}$ be a Banach space representing the state space. Let $\mathcal{PC}\!\para{I, X}$ be the space of piecewise continuous functions from $I$ to $X$, which are right-continuous and the left limit exists for all times $t \in I$. Let the Banach space $\para{U, \norm{\ph}_U}$ represent the input space.
Let furthermore $U_c$ be the space of bounded functions from $I$ to $U$ with norm $\norm{u}_\infty := \sup_{t \in I}\!\braces{\norm{u(t)}_U}$.
We denote the left limit of a function $f$ at $t$ as $f^-(t)$. We consider the standard classes of \textit{comparison functions} $\mathcal{K}$, $\mathcal{K}_\infty$ and $\mathcal{KL}$ as defined by \cite{Sontag1989} and the positive definite function class $\mathcal{P}$ consisting of continuous functions $\gamma:[0,\infty)\to[0,\infty)$, which satisfy $\gamma(0)=0$ and $\gamma(r)>0$ for all $r>0$.
For a continuous function $V: C \to \R$, $C \subset \R$ we denote the \emph{Dini-derivative} by
\begin{align*}
	&\tfrac{\diff}{\diff t}V(t)
	= \limsup\limits_{s \searrow 0} \tfrac{1}{s}\para{V\!\para{s} - V(t)}.
\end{align*}

We consider an impulsive system described by the interacting continuous and discontinuous evolution maps:
\begin{align}\label{eq:ImpulsiveSystem}
    \begin{aligned}
        \dot{x}(t) &= Ax(t) + f\!\para{t,x(t),u(t)}, \quad &&t \in I\setminus S, \\
        x(t) &=g_i\!\para{x^-(t),u^-(t)}, \quad &&t =t_i \in S, \ i \in \N ,
    \end{aligned}
\end{align}
where $u \in U_c$ and $x: I \to X$. The closed linear operator $A: D(A) \to X$ is the infinitesimal generator of a $C_0$-semigroup $T(t)$ on $X$, where $D(A)$ is a dense subset of $X$, $f: I \times X \times U \rightarrow X$, and $g_i: X \times U \rightarrow X$ for all $i \in \N$. We are interested in solutions in the mild sense, i.e., a function $x \in \mathcal{PC}\!\para{I, X}$ such that
\begin{align}\label{eq:ImpulsiveSystemIntegralForm}
    &x(t) = T(t - t_0) x_0 + \int_{t_0}^t T(t - s) f\!\para{s,x(s),u(s)} \,\diff s \nonumber\\
    &+\sum\nolimits_{i \in \N:\ t_i \leq t}T(t - t_i)\! 
    \para{g_i\!\para{x^-(t_i),u^-(t_i)}\! - x^-(t_i)}
\end{align}
holds for all $t \in [t_0, \infty)$ \cite[cf.][]{Ahmed2003a}. We assume that for system \eqref{eq:ImpulsiveSystem}, a (forward-)unique global mild solution exists for every initial condition $x(t_0) = x_0$ and every $u \in U_c$. We denote the value of the solution trajectory at time $t$ with the initial condition $x(t_0) = x_0$ and  the input $u \in U_c$ by $x(t;t_0,x_0,u)$. We shorten the notation by $x(t)$ if the parameters are clear from the context or can be chosen arbitrarily. 

We next define the notion of input-to-state stability.
\begin{definition}\label{def:ISS}
	For a given sequence of impulse times $S$, we call system \eqref{eq:ImpulsiveSystem} \emph{input-to-state stable (ISS)} if there exist functions $\beta \in \mathcal{KL}$ and $\gamma \in \mathcal{K}_\infty$ such that for all initial values $(t_*,x_*) \in I \times X$  and every input function $u \in U_c$, the system has a global solution, which satisfies for all $t \in [t_*, \infty)$
	\begin{align}\label{ineq:ISS}
		\norm{x(t;t_*,x_*,u)}_X \leq \beta\!\para{\norm{x_*}_X, t - t_*} + \gamma\!\para{\norm{u}_\infty}.
	\end{align}
%	\task{We call system \eqref{eq:ImpulsiveSystem} \emph{locally input-to-state stable (LISS)} if there exists a constant $r > 0$ such that \eqref{ineq:ISS} only holds for all $x \in \ball_X(r)$ and all $u \in \ball_{U_c}(r)$.}
\end{definition}

We recall the notion of a candidate ISS-Lyapunov function from \cite{Dashkovskiy2013}.
\begin{definition}\label{def:candLyapunovFunction} Let 
$\Vcand: X \to \R_0^+$ be such that $\Vcand \in \Cont\!\para{X,\R_0^+}$ holds. We call  $\Vcand$ a \emph{candidate ISS-Lyapunov function} for system \eqref{eq:ImpulsiveSystem} if it fulfills  the following conditions:
	\begin{enumerate}[(i)]
		\item There exist functions $\psi_1, \psi_2 \in \mathcal{K}_\infty$ such that
		\begin{align*}
			\psi_1\!\para{\norm{x}_X} \leq \Vcand(x) \leq \psi_2\!\para{\norm{x}_X} %\label{ineq:candLyapunovDefiniteness}
		\end{align*}
		hold true for all $x \in X$. \label{def:candLyapFuncCondition1}
		\item \label{def:candLyapFuncCondition2} There exist functions $\eta \in \mathcal{K}_\infty$ and $\alpha \in \mathcal{P}$ and a continuous function $\rho:\R^+_0 \to \R$ for which $\rho(x) = 0$ $\Leftrightarrow$ $x = 0$ such that for all inputs $u \in U_c$ and all solutions $x = x(t; t_0,x_0,u)$ of \eqref{eq:ImpulsiveSystem}, whenever $\Vcand(x) \geq \eta\!\para{\norm{u}_\infty}$, the inequalities
		\begin{align}
			\tfrac{\diff}{\diff t}\Vcand(x) &\leq - \rho\!\para{\Vcand(x)}, \quad &&t \in I \setminus S, \label{ineq:candLyapunovFlow} \\
			\Vcand\!\para{g_i\!\para{x, u}} &\leq \alpha\!\para{\Vcand(x)}, \quad &&t_i \in S, \label{ineq:candLyapunovJump}
		\end{align}
		hold true.
		\item \label{def:candLyapFuncCondition3_1}There exists a function $\alpha_3 \in \mathcal{K}$ such that for all $x \in X$, all $u \in U_c$, and all $i \in \N$, which satisfy $\Vcand\para{x} < \eta\!\para{\norm{u}_\infty}$, the jump inequality satisfies
		\begin{align}
	    	\Vcand\!\para{g_i\!\para{x, u}} &\leq \psi_3\!\para{\norm{u}_\infty}. \label{ineq:candLyapFuncCondition3}
		\end{align}
	\end{enumerate}
%	\task{We call $\Vcand$ a \emph{candidate LISS-Lyapunov function} for \eqref{eq:ImpulsiveSystem} if there exists a constant $r > 0 $ such that \eqref{ineq:candLyapunovFlow}, \eqref{ineq:candLyapunovJump} and \eqref{ineq:candLyapFuncCondition3} hold for all $x \in \ball_X(r)$ and all $u \in \ball_{U_c}(r)$.}
\end{definition}

\begin{remark}
     Condition \ref{def:LyapFuncCondition3_1} is necessary for establishing ISS, and it was not taken care of in  \cite{Dashkovskiy2013}. Once the state has reached the perturbation radius $\chi\!\para{\norm{u}_\infty}$, it can escape out of it afterward if Condition \ref{def:LyapFuncCondition3_1} is not fulfilled. Consequently, the system will not be ISS. For discrete-time systems, a similar issue was discussed by \cite{Gruene2014}.
\end{remark}

We now provide the notion of time-varying ISS-Lyapunov functions introduced in \cite{Bachmann2022}.

\begin{definition}\label{def:LyapunovFunction}
	Let $V: I \times X \to \R_0^+$ be a function such that $V \in \Cont\!\para{(I\setminus S)\times X,\R_0^+}$ and $V \in \PC(I \times X,\R_0^+)$ hold. We call  $V$ an \emph{ISS-Lyapunov function} for system \eqref{eq:ImpulsiveSystem} if it fulfills all of the following conditions:
	\begin{enumerate}[(i)]
		\item There exist functions $\alpha_1, \alpha_2 \in \mathcal{K}_\infty$ such that
		\begin{align}
			\alpha_1\!\para{\norm{x}_X} \leq V(t,x) \leq \alpha_2\!\para{\norm{x}_X} \label{ineq:LyapunovDefiniteness}
		\end{align}
		holds true for all $t \in I$ and all $x \in X$. \label{def:LyapFuncCondition1}
		\item \label{def:LyapFuncCondition2} There exist functions $\chi \in \mathcal{K}_\infty$ and $\varphi \in \mathcal{P}$ %and a continuous function $\varphi: [0,\infty) \rightarrow \R$ with $\varphi(0) = 0$
		such that for all inputs $u \in U_c$ and all solutions $x = x(t; t_0,x_0,u)$ of \eqref{eq:ImpulsiveSystem}, whenever $V(t, x) \geq \chi\!\para{\norm{u}_\infty}$, the inequalities
		\begin{align}
			\tfrac{\diff}{\diff t}V(t, x) &\leq - \varphi\!\para{V(t,x)}, \quad &&t \in I \setminus S, \label{ineq:LyapunovFlow} \\
			V\!\para{t_i, g_i\!\para{x, u}} &\leq V(t_i^-,x), \quad &&t_i \in S, \label{ineq:LyapunovJump}
		\end{align}
		%holds true for every $t \in I \setminus S$.
		hold true.
		\item \label{def:LyapFuncCondition3_1}There exists a function $\alpha_3 \in \mathcal{K}$ such that for all $x \in X$, all $u \in U_c$, and all $i \in \N$, which satisfy $V\para{t_i^-,x} < \chi\!\para{\norm{u}_\infty}$, the jump inequality satisfies
		\begin{align}
	    	V\!\para{t_i,g_i\!\para{x, u}} &\leq \alpha_3\!\para{\norm{u}_\infty}. \label{ineq:LyapFuncCondition3}
		\end{align}
	\end{enumerate}
%	\task{We call $V$ a \emph{LISS-Lyapunov function} for \eqref{eq:ImpulsiveSystem} if there exists a constant $r > 0 $ such that \eqref{ineq:LyapunovFlow}, \eqref{ineq:LyapunovJump} and \eqref{ineq:LyapFuncCondition3} hold for all $x \in \ball_X(r)$ and all $u \in \ball_{U_c}(r)$.}
\end{definition}

\begin{remark}\label{rem:differenceLFandCLF}
     From inequalities \eqref{ineq:LyapunovFlow} and \eqref{ineq:LyapunovJump}, it follows that the Lyapunov value of a trajectory of the ISS-Lyapunov function strictly falls with time when the state is outside the perturbation radius $\chi\!\para{\norm{u}_\infty}$, whereas the candidate ISS-Lyapunov functions given in Definition \ref{def:candLyapunovFunction} do not strictly fall and additional dwell-time conditions have to be imposed (see Theorems \ref{thm:ISSforSFUJ} and \ref{thm:ISSforUFSJ}) to conclude ISS such that the stabilizing dynamics is dominant over the destabilizing one, which is a restriction for establishing ISS of certain classes of impulsive systems with simultaneous instability in jump and flow.
\end{remark}

\begin{remark}\label{rem:propertiesLF}
    Note that inequality \eqref{ineq:LyapunovFlow} does not necessarily imply that the flow is stable because \eqref{ineq:LyapunovFlow} only holds for time intervals $[t_i,t_{i + 1})$, $i \in \N_0$. In general, $V$ is not continuous and may increase at time instants $t = t_i$, $i \in \N_0$. Similar arguments also hold for inequality \eqref{ineq:LyapunovJump}.
    %\task{Note that the candidate ISS-Lyapunov functions defined by \cite{Dashkovskiy2013} do not strictly fall and additional dwell-time conditions have to be imposed to conclude ISS such that the stabilizing dynamics is dominant over the destabilizing one, which is a restriction for establishing ISS of certain classes of impulsive systems.}
\end{remark}

We define a feedback system related to system~\eqref{eq:ImpulsiveSystem} as follows. Let there exist a function $\eta: X \to \R^+_0$, which is Lipschitz-continuous on bounded subsets of $X$, and a function $\psi \in \mathcal{K}_\infty$ such that $\eta(x) \geq \psi(\norm{x}_X)$ and
\begin{align}
	&\begin{aligned}\label{eq:feedbackImpulsiveSystem}
		\dot{x}(t) &= Ax(t) + f\!\para{t,x(t),d(t)\eta\!\para{x(t)}} \\
		&=:  Ax(t) + \overline{f}\!\para{t,x(t),d(t)}, \quad &t \in I\setminus S, \\
		x(t) &=g_i\!\para{x^-(t),d^-(t)\eta\!\para{x^-(t)}} \\
		&=: \overline{g}_i\!\para{x^-(t),d^-(t)}, \quad &\hspace{-2em} t =t_i, \ i \in \N ,
	\end{aligned}
\end{align}
where $d \in D := \braces{d: I \to U \,\middle|\, \norm{d}_\infty\leq 1}$. We denote trajectories of \eqref{eq:feedbackImpulsiveSystem} with initial condition $x(t_0) = x_0$ by $\overline{x}_d(t; t_0, x_0)$ which we abbreviate by $\overline{x}_d(t)$ if appropriate.

Next, we provide several definitions pertaining to 	system~\eqref{eq:feedbackImpulsiveSystem}.
\begin{definition}\label{def:StabilityDefinitions}
	System \eqref{eq:feedbackImpulsiveSystem} is called
	\begin{enumerate}[(i)]
		\item \emph{uniformly globally asymptotically stable (UGAS)} if there exists a $\mathcal{KL}$-function $\beta$ such that for all $d \in D$, the inequality $\norm{\overline{x}_d(t; t_0,x_0}_X \leq \beta\!\para{\norm{x_0}_X\!, t - t_0}$ holds for all $x_0 \in X$ and all $t \geq t_0$;
		\item \emph{uniformly globally attractive} if for any $C,\eps > 0$ there is a $\delta = \delta(C,\eps) > 0$ such that for all $d \in D$ the inequality $\norm{\overline{x}_d(t; t_0,x_0)}_X \leq \eps$ holds for all $x_0 \in X$ satisfying $\norm{x_0}_X \leq C$ and all $t \geq t_0 + \delta(C,\eps)$;
		\item \emph{uniformly globally stable} if there exists a class $\mathcal{K}_\infty$-function $\alpha$ such that for all $d \in D$, $x_0 \in X$ and $t \geq t_0$ the inequality $\norm{\overline{x}_d(t; t_0,x_0)}_X \leq \alpha\!\para{\norm{x_0}_X}$ holds;
		\item \label{def:robustForwardCompleteness}\emph{robustly forward complete} if for all $C > 0$ and $\tau > t_0$
		\begin{align*}
			K(C,\tau) := \sup\nolimits_{x_0 \in \ball_X(C),\, d \in D,\, t \in [t_0,\tau]} \norm{\overline{x}_d(t;t_0,x_0)}_X
		\end{align*}
		is finite.
	\end{enumerate}
\end{definition}

\begin{definition}\label{def:WURS}
	System \eqref{eq:ImpulsiveSystem} is called \emph{weakly uniformly robustly asymptotically stable (WURS)}, if there exist a function $\eta: X \to \R^+_0$, which is Lipschitz-continuous on bounded subsets of $X$ and $\psi \in \mathcal{K}_\infty$ such that $\eta(x) \geq \psi(\norm{x}_X)$, and  feedback system \eqref{eq:feedbackImpulsiveSystem} is UGAS for all $d \in D$.
\end{definition}

We define a UGAS-Lyapunov function to establish UGAS of feedback system \eqref{eq:feedbackImpulsiveSystem}. Note that it is not an ISS-Lyapunov function as given in Definition~\ref{def:LyapunovFunction} but it provides global asymptotic stability, uniformly with respect to control.
\begin{definition}\label{def:LyapunovFunctionFeedback}
    Let $V: I \times X \to \R_0^+$ be such that $V \in \Cont\!\para{(I\setminus S)\times X,\R_0^+}$ and $V \in \PC(I \times X,\R_0^+)$ hold. We call  $V$ a \emph{UGAS-Lyapunov function} for  system \eqref{eq:feedbackImpulsiveSystem}, if it fulfills all of the following conditions:
	\begin{enumerate}[(i)]
		\item There exist functions $\alpha_1, \alpha_2 \in \mathcal{K}_\infty$ such that
		\begin{align*}
			\alpha_1\!\para{\norm{\overline{x}}_X} \leq V(t,\overline{x}) \leq \alpha_2\!\para{\norm{\overline{x}}_X} %\label{ineq:LyapunovDefiniteness}
		\end{align*}
		holds true for all $t \in I$ and all $\overline{x} \in X$. %\label{def:LyapFuncCondition1}
		\item There exists a $\varphi \in \mathcal{P}$ %and a continuous function $\varphi: [0,\infty) \rightarrow \R$ with $\varphi(0) = 0$
		such that for all $u \in U_c$ and all $\overline{x} = \overline{x}(t; t_0,x_0,u)$ of \eqref{eq:ImpulsiveSystem}, the inequalities
		\begin{align*}
			\tfrac{\diff}{\diff t}V(t, \overline{x}) &\leq - \varphi\!\para{V(t,\overline{x})}, \quad &&t \in I \setminus S, %\label{ineq:LyapunovFlow} 
			\\
			V\!\para{t_i, \overline{g}_i\!\para{\overline{x}, u}} &\leq V(t_i^-,\overline{x}), \quad &&t_i \in S, %\label{ineq:LyapunovJump}
		\end{align*}
		hold true.
	\end{enumerate}
\end{definition}

%\begin{remark}
%    We show a method on constructing an ISS-Lyapunov function from a candidate ISS-Lyapunov function in Appendix \ref{app:constructionLF}.
%\end{remark}

%\input{chapters/02b_example_superiority}

% !TeX spellcheck = en_US
\section{Necessary and sufficient condition for ISS}

In this section, we recall a necessary and sufficient condition for ISS from our recent work \cite{Bachmann2022} for the sake of completeness. But before formally stating and proving the main result, we provide an assumption.
\begin{assumption}\label{A1}
	Let $f: I \times X \times U \to X$ be Lipschitz continuous in the second and third variable on bounded sets, i.e., for all $C,D > 0$, there exist constants $L_f^1(C,D), L_f^2(C,D) > 0$ such that for all $x,y \in \ball_X(C)$ and all $u, v \in \ball_U(D)$ the  inequalities 
	\begin{enumerate}[(i)]
		\item \label{A1:first} $\norm{f(t,x,u) - f(t,y,u)}_X \leq L_f^1(C,D)\norm{x - y}_X$,
		\item \label{A1:second} $\norm{f(t,x,u) - f(t,x,v)}_X \leq L_f^2(C,D)\norm{u - v}_U$,
	\end{enumerate}
	hold uniformly in time $t \in I$. Furthermore, let $f$ be continuous in the first variable for all $t \in I\setminus S$.
	
	Let $g_i: X \times U \to X$ for all $i \in \N$ be Lipschitz continuous in the first and second variable on bounded sets, i.e. for all $C,D > 0$, there exist constants $L_{g_i}^1(C,D),\, L_{g_i}^2(C,D) > 0$ such that for all $x,y \in \ball_X(C)$ and all $u, v \in \ball_U(D)$ the inequalities 
	\begin{enumerate}[(i)]
		\item[(iii)] $\norm{g_i(x,u) - g_i(y,u)}_X \leq L_{g_i}^1(C,D)\norm{x - y}_X$,
		\item[(iv)] $\norm{g_i(x,u) - g_i(x,v)}_X \leq L_{g_i}^2(C,D)\norm{u - v}_U$,
	\end{enumerate}
	hold.
\end{assumption}

\begin{theorem}\label{thm:Lyapunov}
	Let there exist an ISS-Lyapunov function for impulsive system \eqref{eq:ImpulsiveSystem}, then it is ISS.
\end{theorem}

\begin{pf}
    We define the set 
    \begin{align*}
    	A_1(t) := \braces{x \in X \, \middle| \, V(t,x) < \chi\!\para{\norm{u}_{\infty}}}.
    \end{align*}
    At first, we show that for all initial conditions $(x_0, t_0)$ for which $x_0 \notin A_1(t_0)$, there exists a $\mathcal{KL}$-function $\beta$ such that the inequality
    \begin{align*}
        {\norm{x(t;t_0,x_0,u)}_X \leq \beta\!\para{\norm{x_0}_X\!, t - t_0}}
    \end{align*}
    holds true as long as $x(t_*) \notin A_1(t_*)$ for all $t_* \in [t_0,t]$. Afterwards, we show that trajectories, once they have reached the set $A_1$, will stay bounded.
    
    Step 1: Let $x(t) \notin A_1$, i.e. $\V \geq \chi\!\para{\norm{u}_\infty}$ for all $(t,x,u)$ such that \eqref{ineq:LyapunovFlow} and \eqref{ineq:LyapunovJump} are satisfied.
    For brevity, we use $v: I \to \R^+_0,\, v(t) := V\!\para{t,x(t)}$ and denote $v_i := v(t_i)$ and $v^-_i := v^-(t_i)$.
    For $v(t) \neq 0$, \eqref{ineq:LyapunovFlow} becomes
    \begin{align}\label{ineq:transformIDEforV_flow}
    	\frac{\frac{\diff}{\diff t}v}{\varphi(v)}
    	\leq\frac{-\varphi(v)}{\varphi(v)} = -1.
    \end{align}
    Note that the case $v(t^*) = 0$ for some $t^* \geq t_0$ implies $v \equiv 0$ on $[t^*, \infty)$.
    
    Integrating \eqref{ineq:transformIDEforV_flow} over the interval $[t_i,t_*]$, $i \in \N_0$, for some $t_* \in [t_i,t_{i+1})$, we obtain
    \begin{align}\label{ineq:estimateFlowVariableUpperBound}
        \int_{v_i}^{v(t_*)} \frac{1}{\varphi(s)}\ \diff s
    	= \int_{t_i}^{t_*} \frac{\frac{\diff}{\diff t} v(t)}{\varphi\!\para{v(t)}}\ \diff t\leq -(t_* - t_i),
    \end{align}
    where $s := v(t)$.
    
    We define the function $F: [0,\infty) \rightarrow \R \cup \{-\infty\}$ as
    \begin{align*}
    	F(q) := \int_{1}^{q} \frac{1}{\varphi(s)} \ \diff s,
    \end{align*}
    where $F$ is strictly increasing such that the inverse $F^{-1}: \image(F) \rightarrow [0,\infty)$ exists.
    
    %%%%%%%%%%%%%%%%%%%%%%%%%%%%%%%%%%%%%%%%%%%%%%%%%%%%%%%%%%%%%%%%%%%%%%%
    % find estimate for average impulsive behavior
    Let $n = n(t) := \max\!\braces{n \in \N\,\middle|\, t_n \leq t}$.
    By monotonicity of $F$ and \eqref{ineq:LyapunovJump}, we have $F(v_i) - F(v^-_i)\leq 0$ for all $i \in \N$. Combining this and substituting $F$ in  \eqref{ineq:estimateFlowVariableUpperBound}, we obtain
    \begin{align}
    	&F(v(t)) - F(v_0) \nonumber\\
    	& = F(v(t))\! -\! F(v_n) \!+ \!\sum\limits_{k = 1}^n F(v_k) \! - \! F(v^-_k) \! + \!
    	F(v^-_k) \! - \! F(v_{k - 1}) \nonumber\\
    	&\leq - (t - t_n) -\sum\nolimits_{k = 1}^n t_k - t_{k - 1}
    	= -(t - t_0). \label{ineq:estimateV}
    \end{align}
    This is equivalent to
    \begin{align}\label{ineq:vinF}
    	v(t) \leq F^{-1}\!\para{F\!\para{v_0} - (t - t_0)}.
    \end{align}
    Due to the case distinction in case $v(t_*) = 0$ for some $t_* \in I$ mentioned above it is appropriate to choose a more relaxed estimate than \eqref{ineq:vinF} by
    \begin{align*}
        \tilde{\beta}\!\para{v_0, t - t_0}\!&:= F^{-1}\!\!\left(\!F\!\para{v_0} - \para{F\!\para{v_0} - m}\!\para{\!1-\displaystyle e^{\frac{-(t - t_0)}{F\!\para{v_0} - m}}\!}\right) \\
        &\geq F^{-1}\!\para{F\!\para{v_0} - (t - t_0)}
    \end{align*}
    for $m > - \infty$, where $m := F^{-1}(0)$. Otherwise, we set 
    \begin{align*}
        \tilde{\beta}\!\para{v_0, t - t_0} := F^{-1}\!\para{F\!\para{v_0} - (t - t_0)}.
    \end{align*}
    In both cases we have $\tilde{\beta} \in \mathcal{KL}$ and we define $\beta \in \mathcal{KL}$, $\beta(r,s) := \alpha_1^{-1}\!\para{\tilde{\beta}\!\para{\alpha_2(r),s}}$ to obtain the desired bound
    \begin{align} \label{ineq:asymptoticStabilityOfTrajectory}
    	\norm{x(t;t_0,x_0,u)}_X \leq \beta\!\para{\norm{x_0}_X\!, t - t_0}
    \end{align}
    for $t \in [t_0, t^*]$, $t^* := \inf\!\braces{t \in [t_0,\infty] \, \middle| \, \V\! <\! \chi\!\para{\norm{u}_{\infty}}}$.
    
    Step 2: Next, we  show that trajectories that are in $A_1(t)$ for some $t \in I$, stay bounded for all times. Therefore, we apply \eqref{ineq:LyapFuncCondition3} from which we can conclude that all trajectories jumping from $A_1^-(t)$ for some $t = t_i$ are bounded by
    \begin{align*}
    	A_2(t) := \braces{x \in X \,\middle| \, V(t,x) \leq \alpha_3\!\para{\norm{u}_\infty}}.
    \end{align*}
    It is not possible that the trajectories leave
    \begin{align*}
    	&A_3(t) := \overline{A_1(t) \cup A_2(t)} \\
    	&= \braces{x \in X \,\middle| \, V(t,x) \leq \max\!\braces{\alpha_3\!\para{\norm{u}_\infty}\!, \chi\!\para{\norm{u}_{\infty}}}}
    \end{align*}
    neither by jump nor by flow because the boundary of $A_1(t) \cup A_2(t)$ is in the complement of $A_1$. From \eqref{ineq:LyapunovFlow}, it follows that $\frac{\diff}{\diff t}\V < 0$ holds on the boundary, and jump inequality \eqref{ineq:LyapunovJump} prevents the trajectories from leaving $A_3(t)$. We define $\gamma \in \mathcal{K}_\infty$, $\gamma :=\alpha_1^{-1} \circ \max\!\braces{\alpha_3\!\para{\ph}, \chi\!\para{\ph}}$. Then $\norm{x(t;t_0,x_0,u)}_X \leq \gamma\!\para{\norm{u}_\infty}$
    holds for all $t > t^*$.
    From this equation and \eqref{ineq:asymptoticStabilityOfTrajectory}, we can conclude
    \begin{equation*}
    	\norm{x(t;t_0,x_0,u)}_X \leq \beta\para{\norm{x_0}_X\!, t - t_0} + \gamma\!\para{\norm{u}_\infty}.
    \end{equation*}
\end{pf}

\begin{theorem}\label{thm:converseLyapunov}
	Let impulsive system \eqref{eq:ImpulsiveSystem} be ISS and satisfy Assumption 1. Then, there exists an ISS-Lyapunov function for system \eqref{eq:ImpulsiveSystem}.
\end{theorem}

% \begin{theorem}\label{thm:existenceOfLyapunovFunction}
%     Let system (1) be WURS, and  its solutions be Lipschitz continuous with respect to initial values in bounded subsets of $X$ and locally Lipschitz continuous on intervals $t \in [t_i,t_{i + 1})$, $i \in N_0$, where the Lipschitz constant depends on the norm of the initial condition $\norm{x_0}_X$. Then, there is a Lyapunov function $V$ for system \eqref{eq:feedbackImpulsiveSystem} as given in Definition \ref{def:LyapunovFunctionFeedback}, which is Lipschitz continuous on bounded balls in space and locally Lipschitz continuous on the intervals $[t_i,t_{i + 1})$, $i \in \N_0$.
% \end{theorem}

\begin{pf}
	Let system \eqref{eq:ImpulsiveSystem} be ISS. By Lemma \ref{lem:WURS} in the Appendix, it is WURS.
	Therefore, feedback system \eqref{eq:feedbackImpulsiveSystem} is UGAS for all $d \in D$ with  $\eta$ and $\psi$ as given in Definition~\ref{def:WURS}. Proposition \ref{prop:Lipschitzcontinuity} in the Appendix then gives us that every solution of system \eqref{eq:feedbackImpulsiveSystem} is Lipschitz continuous with respect to initial values in bounded subsets of $X$, and locally Lipschitz continuous on the intervals $[t_i,t_{i + 1})$, $i \in \N_0$. With Proposition \ref{prop:Lipschitzcontinuity} at hand, the converse Lyapunov theorem by \citet[Thm. 3.4]{Karafyllis2011} ensures that there exists a UGAS-Lyapunov function $V: I \times X \to \R^+_0$ for  feedback system \eqref{eq:feedbackImpulsiveSystem}, which is Lipschitz continuous on bounded balls in space and locally Lipschitz continuous on the intervals $[t_i,t_{i + 1})$, $i \in \N_0$. From Lemma \ref{lem:WURStoISS} in the Appendix, it follows that there exists an ISS-Lyapunov function for system \eqref{eq:ImpulsiveSystem} as stated in the theorem.
\end{pf}

\section{Main results}\label{sec:constructionLF}

As the main concern of this paper, we study the construction of an ISS-Lyapunov function from a candidate ISS-Lyapunov function. More precisely, we show a method for constructing an ISS-Lyapunov function $V = V(t,x)$ as in Definition \ref{def:LyapunovFunction} from a candidate ISS-Lyapunov function $\Vcand$ with rates $\varphi$ and $\alpha$ as defined in Definition \ref{def:candLyapunovFunction}. 
\subsection{Stable flows and unstable jumps}

Let us consider system~\eqref{eq:ImpulsiveSystem} with stable flows and unstable jumps. Theorem~1 of \cite{Dashkovskiy2013} provides the following sufficient stability result.
\begin{theorem}\label{thm:ISSforSFUJ}
    Let $\Vcand$ be a candidate ISS-Lyapunov function for system \eqref{eq:ImpulsiveSystem} and $\rho, \alpha$ be as in Definition~\ref{def:candLyapunovFunction} and $\rho \in \mathcal P$. If for some $\theta, \delta > 0$ and all $a > 0$ it holds that
    \begin{align}\label{ineq:dwellTimeCondition}
        \int_a^{\alpha(a)}\frac{1}{\rho(s)} \,\diff s \leq \theta - \delta
    \end{align}
    then \eqref{eq:ImpulsiveSystem} is ISS for all impulse time sequences $S$ for which $\theta \leq \inf_{i \in \N} \{t_{i + 1} - t_i\}$ holds.
\end{theorem}
Now we show that ISS-Lyapunov functions  given in Definition \ref{def:LyapunovFunction} can be constructed from $\Vcand$ given in the previous theorem.

\begin{theorem}
    Let $\Vcand$ be a candidate ISS-Lyapunov function for system \eqref{eq:ImpulsiveSystem} and parameters $\rho, \alpha, \theta$ and $\delta$, as specified in Theorem~\ref{thm:ISSforSFUJ}, fulfill \eqref{ineq:dwellTimeCondition}. Let $\tilde{F}(q) :=  \int_1^q \frac{1}{\rho(s)}\,\diff s$.
    Let $\kappa \in \mathcal K_\infty \cap \Cont^1(\R^+_0, \R^+_0)$ such that $\kappa \leq \min\!\braces{\alpha^{-1}, \operatorname{id}}$ and $\kappa' \in \mathcal P$ hold. Then, an ISS-Lyapunov function is given by
    \begin{align*}
        V(t,x) := \max\!\braces{v_1(t,x), v_2(t,x)}
    \end{align*}
    where $v_1, v_2: I \times X \to \R^+_0$ are given by
    \begin{align*}
        v_1(t,x) 
        &:= \tilde F^{-1}\!\para{\!\max\!\braces{\!\tilde F\!\para{\Vcand(x)} - \tfrac{t_{i + 1} - t}{t_{i + 1} - t_i}\!\para{\theta \! - \! \delta}, \tilde F(0)\!}} \\
        v_2(t,x)
        &:= \kappa\!\para{\Vcand(x)}
    \end{align*}
    for $t \in [t_i, t_{i+1})$, $i \in \N$.
\end{theorem}
 
The role of $v_2$ is to prevent possible issues with finite time convergence. In that case, $F\!\para{\Vcand(x)} - \tfrac{t_{i + 1} - t}{t_{i + 1} - t_i}\para{\theta - \delta}$ might become smaller than zero. Then $v_1(t,x)$ is not strictly positive for $x \neq 0$ anymore.

\begin{pf}
    Note that $\tilde F$ is strictly increasing and invertible (cf. $F$ in the proof of Theorem \ref{thm:Lyapunov}).
    
    We show that all the conditions of Definition \ref{def:LyapunovFunction} are fulfilled.
    \begin{enumerate}
        \item By the definition of $\kappa$, it follows that
        \begin{align*}
            V(t,x) 
            &\geq \kappa\!\para{\Vcand(x)} \geq \kappa\!\para{\psi_1\!\para{\norm{x}_X}}
            =: \alpha_1\!\para{\norm{x}_X}.
        \end{align*}
        Analogously, we obtain
        \begin{align*}
            V(t,x) 
            &\leq \max\!\braces{\Vcand(x), \kappa\!\para{\Vcand(x)}} \\
            &=\Vcand(x)
            \leq \psi_2\!\para{\norm{x}_X}
            =: \alpha_2\!\para{\norm{x}_X}.
        \end{align*}
        \item Next, we treat $V(t,x) \geq \chi\!\para{\norm{u}_\infty} := \alpha\!\para{\eta\!\para{\norm{u}_\infty}}$. Then $\Vcand(x) \geq \alpha^{-1}\!\para{V(t,x)} \geq\eta\!\para{\norm{u}_\infty}$. We bound the Dini-derivative 
        \begin{align}
            &\frac{\diff}{\diff t} V(t,x) \nonumber\\
            &\leq
            \begin{cases}
                \dot v_1(t,x), \quad &\text{if } v_1(t,x) > v_2(t,x), \\
                \dot v_2(t,x), \quad &\text{if } v_1(t,x) < v_2(t,x), \\
                \max\!\braces{\dot v_1(t,x), \dot v_2(t,x)}, &\text{if } v_1(t,x) = v_2(t,x).
            \end{cases}
            \label{ineq:boundV}
        \end{align}
        For $v_1(t,x) \geq v_2(t,x)\geq 0$, we have
        \begin{align}
            &\dot v_1(t,x)\\
            &\leq \para{\tilde F^{-1}}'\!\para{\tilde F\!\para{\Vcand(x)} - \tfrac{t_{i + 1} - t}{t_{i + 1} - t_i}\para{\theta - \delta}} \nonumber\\*
            &\hspace{-18mm}\hphantom{\dot v_1(t,x)
            \leq}\qquad \times \para{\tfrac{\dot V_{\operatorname{cand}}(x)}{\rho\para{\Vcand(x)}} + 1 - \tfrac{\delta}{\theta}} \nonumber\\
            &\leq - \frac{\delta}{\theta} \ \rho\!\para{\tilde F^{-1}\!\para{\tilde F\!\para{\Vcand(x)} -  \tfrac{t_{i + 1} - t}{t_{i + 1} - t_i}\para{\theta - \delta}}} \nonumber\\
            &= - \frac{\delta}{\theta} \ \rho\!\para{v_1(t,x)}, \label{ineq:boundv1}
        \end{align}
        where we have used $t_{i + 1} - t_i \geq \theta$ and the definition of $\tilde F$ in the first step and the inverse function theorem in the second step. Furthermore,
        \begin{align}
            \dot v_2(t,x)
            &= \kappa'\!\para{\Vcand(x)}\dot V_{\operatorname{cand}}(x) \nonumber\\
            &\leq -\kappa'\!\para{\Vcand(x)}\rho\!\para{\Vcand(x)} \label{ineq:boundv2}
        \end{align}
        holds.
        As $\rho, \kappa' \in \mathcal P$ there exists such a function $\varphi \in \mathcal P$ that from
        \eqref{ineq:boundV}, \eqref{ineq:boundv1}, and \eqref{ineq:boundv2} follows.
        \begin{align*}
            \frac{\diff}{\diff t} V(t,x)
            &\leq -\varphi(V(t,x))
        \end{align*}
        Therefore, $V(t,x)$ is strictly falling. For the jumps, we find the bound
        \begin{align}
            &V(t_i,g_i(x,u)) \nonumber\\
            &\!\leq\! \max\!\braces{\!\tilde F^{-1}\!\!\para{\!\tilde F\!\para{\alpha\!\para{\Vcand(x)}} \!-\! \para{\theta \!-\! \delta}\!}\!\!,\! \kappa\!\para{\alpha\!\para{\Vcand(x)}}\!} \nonumber\\
            &\!\leq\! \max\!\braces{\tilde F^{-1}\!\para{\tilde F\!\para{\Vcand(x)}}\!, \Vcand(x)} \nonumber\\
            &\!\leq\! \tilde F^{-1}\!\para{\tilde F\!\para{\Vcand(x)} - \tfrac{t_i - t_i}{t_i - t_{i - 1}}\para{\theta - \delta}} \nonumber\\
            &\!\leq\! V^-(t_i,x), \label{ineq:LyapunovJumpEstimate}
        \end{align}
        where we used \eqref{ineq:dwellTimeCondition} in the second inequality.
        \item For $V^-(t_i,x) < \chi\!\para{\norm{u}_\infty}$ we differ the two cases $\Vcand(x) < \eta\!\para{\norm{u}_\infty}$ for which
        \begin{align*}
            V(t_i,g_i(x,u))
            &\leq \Vcand(g_i(x,u)) 
            \leq \psi_3\!\para{\norm{u}_\infty}
        \end{align*}
        holds true and $\Vcand(x) \geq \eta\!\para{\norm{u}_\infty}$ for which follows by the same steps as in \eqref{ineq:LyapunovJumpEstimate}
        \begin{align*}
            V(t_i,g_i(x,u))
            &\leq V^-(t_i,x) \leq \chi\!\para{\norm{u}_\infty}.
        \end{align*}
        Hence, we choose $\alpha_3 \in \mathcal K$, $\alpha_3(a) := \max\{\psi_3(a),\chi(a) \}$.
    \end{enumerate}
    By this, all conditions of an ISS-Lyapunov function are fulfilled.
\end{pf}

\subsection{Unstable flows and stable jumps}
Let us consider system~\eqref{eq:ImpulsiveSystem}  with unstable flows and stable jumps. We recall the following sufficient stability result from \cite[Theorem 3]{Dashkovskiy2013}.
\begin{theorem}\label{thm:ISSforUFSJ}
    Let $\Vcand$ be a candidate ISS-Lyapunov function for impulsive system \eqref{eq:ImpulsiveSystem} and $\rho, \alpha$ are as in Definition~\ref{def:candLyapunovFunction} with $-\rho \in \mathcal P$. If for some $\theta, \delta > 0$ and all $a > 0$ it holds that
    \begin{align}\label{ineq:dwellTimeConditionUFSJ}
        \int_{\alpha(a)}^a \frac{1}{-\rho(s)} \,\diff s \geq \theta - \delta
    \end{align}
    then system \eqref{eq:ImpulsiveSystem} is ISS for every sequence $S$ satisfying $\theta \geq \sup_{i \in \N} \{t_{i + 1} - t_i\}$.
\end{theorem}

Now we show that ISS-Lyapunov functions  given in Definition \ref{def:LyapunovFunction} can be constructed from $\Vcand$ given by Theorem~\ref{thm:ISSforUFSJ}.

\begin{theorem}
    Let $\Vcand$ be a candidate ISS-Lyapunov function for system \eqref{eq:ImpulsiveSystem} and parameters $\rho, \alpha, \theta$ and $\delta$ as specified in Theorem~\ref{thm:ISSforUFSJ} fulfill \eqref{ineq:dwellTimeConditionUFSJ}. Let $\tilde{F}(q) :=  \int_1^q \frac{1}{-\rho(s)}\,\diff s$.
    
    Then, an ISS-Lyapunov function is defined by
    \begin{align*}
        V(t,x) := \tilde F^{-1}\!\para{\tilde F\!\para{\Vcand(x)} - \tfrac{t - t_i}{t_{i + 1} - t_i}\para{\theta + \delta}}
    \end{align*}
    for $t \in [t_i, t_{i+1})$, $i \in \N$.
\end{theorem}
The proof follows similar argumentation as the case with stable jumps and unstable flows and therefore is omitted here.

\section{Conclusion}\label{sec:conclusion}
We proposed a method to construct time-varying ISS-Lyapunov functions from candidate ISS Lyapunov functions. By this, we could show that the new concept of time-varying ISS-Lyapunov functions integrates into the existing stability theory for impulsive systems. Even more, time-varying Lyapunov functions can serve as a new standard formulation for Lyapunov methods for impulsive systems as the existence of ISS-Lyapunov functions is guaranteed by Theorem \ref{thm:converseLyapunov}. Moreover, the time-varying Lyapunov functions cover broad system classes (including systems with simultaneous instability in continuous and discrete dynamics), which is a fundamental advantage over candidate ISS-Lyapunov functions.

%\begin{ack}
    %We thank Sergey Dashkovskiy for his useful suggestions and communicating the result of \cite{Dashkovskiy2021a} to us.
%\end{ack}

\appendix

\section{Technical Results}
\begin{lemma}\label{lem:WURS}
	If impulsive system \eqref{eq:ImpulsiveSystem} is ISS, then it is WURS.
\end{lemma}

\begin{pf}
    %	First show: from ISS follows WURS
From the definition of ISS, it follows that there exist functions $\beta \in \mathcal{KL}$ and $\gamma \in \mathcal{K}_\infty$. We define $\alpha(s) := \beta(s,0)$ for all $s \in \R^+_0$. From inequality \eqref{ineq:ISS}, it follows that $\alpha(s) \geq s$ for all $s \in \R^+_0$, which means that $\alpha \in \mathcal{K}_\infty$.

We define $\sigma \in \mathcal{K}_\infty$ as $\sigma(s) \leq \para{\gamma^{-1}\!\para{\tfrac{1}{4} \alpha^{-1}\!\para{\tfrac{2}{3}\alpha^{-1}\!\para{\tfrac{2}{3} s}}} }$, and choose locally Lipschitz continuous functions $\eta: X \to \R^+_0$ and $\psi  \in \mathcal{K}_\infty$  such that  $\psi\!\para{\norm{x}_X} \leq \eta(x) \leq \sigma\!\para{\norm{x}_X}$.

We show that with this definition of $\eta$, the inequality
\begin{align}\label{ineq:estimategammau}
	\gamma\!\para{\norm{d(t)\eta\!\para{\overline{x}_d(t}}_U} \leq \tfrac{1}{2} \norm{x_0}_X
\end{align}
holds for all $t \geq t_0$. We first compute
\begin{align*}
	&\gamma\!\para{\norm{d(t) \eta\!\para{\overline{x}_d(t}}_U} 
	\leq \gamma\!\para{\sigma\!\para{\norm{\overline{x}_d(t)}_X}} \\*
	&\leq \tfrac{1}{4} \alpha^{-1}\!\para{\tfrac{2}{3}\alpha^{-1}\!\para{\tfrac{2}{3}\norm{\overline{x}_d(t)}_X}}
	\leq\tfrac{1}{9}\norm{\overline{x}_d(t)}_X
\end{align*}
which is valid for every $d \in D$, all $x_0 \in X$, and all $t \in I$. By definition, $\overline{x}_d$ is right-continuous, so for each $t_* \in I$, there exists a $\delta > 0$ such that
\begin{align}
	\gamma\!\para{\norm{d(t) \eta\!\para{\overline{x}_d(t;t_*, x_*)}}_U}
	&\!\leq\!\tfrac{1}{9}\!\norm{\overline{x}_d(t; t_*, x_*)}_X\!
	\!\leq\! \tfrac{1}{2}\! \norm{x_*}_X \label{ineq:estimategamma}
\end{align}
for all $t \in [t_*,t_* + \delta]$, where the last inequality in \eqref{ineq:estimategamma} follows from continuity. Next, we fix random $d \in D$, $t_0 \in I$, and $x_0 \in X$, and define $\overline{t} = \overline{t} (t_0, x_0, d)$ by
\begin{align*}
	&\overline{t} = \inf \!\braces{t \geq t_0 \,\middle|\, \gamma\!\para{\norm{d(t) \eta\!\para{\overline{x}_d(t}}_U} > \tfrac{1}{2} \!\norm{x_0}_X}\!.
\end{align*}
We prove by contradiction that $\overline{t} = \infty$. Thus, let us assume that $\overline{t}$ is not equal to infinity. Then, inequality \eqref{ineq:estimategamma} holds for all $t \in [t_0,\overline{t})$. By substituting inequality \eqref{ineq:estimategamma} into \eqref{ineq:ISS},
\begin{align}
	&\norm{\overline{x}_d(t)}_X\! 
	\leq\! \beta\!\para{\norm{x_0}_X\!, t - t_0} + \tfrac{1}{2}\!\norm{x_0}_X \!
	\leq\! \tfrac{3}{2}\alpha\!\para{\norm{x_0}} \label{ineq:estimateTrajectoryByIC}
\end{align}
is obtained. System \eqref{eq:ImpulsiveSystem} follows the principle of causality, i.e., it only depends on inputs $u(t)$ for $t \in [t_0, \overline{t})$.
Therefore, we can transform \eqref{ineq:ISS} into
\begin{align}
	&\norm{\overline{x}_d(\overline{t};t_*,x_*)}_X 
	\leq \beta\!\para{\norm{x_*}_X, \overline{t} - t_*} \nonumber\\
	&+ \gamma\!\para{\sup\nolimits_{t \in [t_*,\overline{t})}\!\braces{\norm{d(t)\eta\!\para{\overline{x}_d(t;t_*, x_*)}}_U}}\!.\label{ineq:ISSmod}
\end{align}
As system \eqref{eq:feedbackImpulsiveSystem} is only right-continuous, we need an estimate for limits from the left, which also considers possible jumps.  This estimate is given by 
\begin{align}
	&\norm{\overline{x}_d(\overline{t})}_X
	= \lim\nolimits_{t_* \nearrow \overline{t}} \norm{\overline{x}_d(\overline{t};t_*,\overline{x}_d(t_*))}_X \nonumber\\
	&\leq \lim\nolimits_{t_* \nearrow \overline{t}}\beta\!\para{\norm{\overline{x}_d(t_*)}_X, \overline{t} - t_*} \nonumber\\
	&\quad + \gamma\!\para{\sup\nolimits_{t \in [t_*,\overline{t})}\!\braces{\norm{d(t) \eta\!\para{\overline{x}_d\!\para{t_*}}}_U}} \nonumber\\
	&\leq \alpha\!\para{\norm{\overline{x}^-_d(\overline{t})}_X} + \tfrac{1}{2}\!\norm{\overline{x}^-_d(\overline{t})}_X 
	\leq \tfrac{3}{2}\alpha\!\para{\norm{\overline{x}^-_d(\overline{t})}_X}\!, \label{ineq:estimatexatjump}
\end{align}
which we get from the semigroup property $\overline{x}_d(\overline{t}) = \overline{x}_d\!\para{\overline{t};t,\overline{x}_d(t)}$ for $t \in [t_0,\overline{t}]$  and \eqref{ineq:ISSmod}.
From \eqref{ineq:estimatexatjump} follows
\begin{align*}
	\gamma\!\para{\norm{d(\overline{t})\eta\!\para{\overline{x}_d(\overline{t})}}_U}
	&\leq \gamma\!\para{\sigma\!\para{\tfrac{3}{2}\alpha\!\para{\norm{\overline{x}^-_d(\overline{t})}_X}}} \\
	&\leq \tfrac{1}{4}\alpha^{-1}\!\para{\tfrac{2}{3} \norm{\overline{x}^-_d(\overline{t})}_X}
	\leq \tfrac{1}{4}\norm{x_0}_X,
\end{align*}
where we have used the definition of $\sigma$ in the second, and \eqref{ineq:estimateTrajectoryByIC} in the third inequality.

As $\overline{x}_d$ is right-continuous, there is a neighborhood to the right of $\overline{t}$ such that $\gamma\!\para{\norm{d(t) \eta\!\para{\overline{x}_d(t}}_U} \leq \tfrac{1}{2} \norm{x_0}_X$. This is a contradiction to the definition of $\overline{t}$ and shows the validity of \eqref{ineq:estimategammau} for all $t \geq t_0$.
Therefore,
\begin{align}\label{ineq:estimatexetainbeta}
	\norm{\overline{x}_d(t)}_X \leq \beta\!\para{\norm{x}_X, t - t_0} + \tfrac{1}{2}\norm{x}_X
\end{align}
holds for all $x \in X$, all $d \in D$, and all $t \geq t_0$, which gives us uniform global stability of system \eqref{eq:feedbackImpulsiveSystem}.

To show uniform global attractivity, we use the fact that there exists such a $\tau_1 =\tau_1\!\para{\norm{x}_X}$ that ${\beta\!\para{\norm{x}_X\!, t} \!\leq\! \tfrac{1}{4}\! \norm{x}_X}$ for all $t \geq \tau_1$. Substituting this term into \eqref{ineq:estimatexetainbeta} yields $\norm{\overline{x}_d(t)}_X \leq \tfrac{3}{4}\norm{x_0}_X$ for all $x_0 \in X$, $d \in D$, and $t \geq \tau_1$. Recursively one can define a sequence $(\tau_i)_{i \in \N}$ that only depends on $\norm{x_0}_X$ such that
\begin{align*}
    \norm{\overline{x}_d(t)}_X \leq \para{\tfrac{3}{4}}^k\norm{x_0}_X
\end{align*}
holds for all $x_0 \in X$, $d \in D$, and $t \geq \tau_k$. This implies uniform global attractivity of system \eqref{eq:feedbackImpulsiveSystem}.
From \citet[Thm. 2.2]{Karafyllis2011} follows UGAS of system \eqref{eq:feedbackImpulsiveSystem}. Hence, impulsive system \eqref{eq:ImpulsiveSystem} is WURS by Definition \ref{def:WURS}.
\end{pf}

%{\color{green}

\begin{lemma}\label{lem:LipschitzcontinuityRHS}
	Let Assumption \ref{A1} hold. Then $\overline{f}$ is Lipschitz continuous in the second variable on bounded subsets of $X$ uniformly with respect to the third argument and for all $t \in I$, i.e., for all $C > 0$, there exists a constant $L_{\overline{f}}(C) > 0$ such that for all $x,y \in \ball_X(C)$ and all $d \in D$, the inequality
	\begin{align*}
		\norm{\overline{f}\!\para{t, x, d} - \overline{f}\!\para{t, y, d}}_X \leq L_{\overline{f}}(C) \norm{x - y}_X
	\end{align*}
	is satisfied.	
	Moreover, if Assumption \ref{A1} holds, $\overline{g}_i$ is Lipschitz in the first variable on bounded subsets of $X$ for every $i \in \N$ uniformly with respect to the second argument, i.e., for all $C > 0$, there exists an $L_{\overline{g}_i}(C) > 0$ such that for all $x,y \in \ball_X(C)$ and all $d \in D$, follows
	\begin{align*}
	\norm{\overline{g}_i\!\para{x, d} - \overline{g}_i\!\para{y, d}}_X \leq L_{\overline{g}_i}(C) \norm{x - y}_X.
	\end{align*}
\end{lemma}
\begin{pf}
	Let us choose $C > 0$ arbitrarily, and fix  $x,y \in \ball_X(C)$ and $d \in D$. By definition, $\eta$ is locally Lipschitz, i.e., there exists a bound $R = R(C) > 0$ such that $\eta(x) \leq R$ for all $\norm{x}_X\leq C$. Then follows
	\begin{align*}
		&\norm{\overline{f}\!\para{t, x, d(t)} - \overline{f}\!\para{t, y, d(t)}}_X \\
		&= \norm{f\!\para{t,x,d(t)\eta(x)} - f\!\para{t,y,d(t)\eta(y)}}_X \\
		&\leq \norm{f\!\para{t,x,d(t)\eta(x)} - f\!\para{t,y,d(t)\eta(x)}}_X \\
		&\quad+ \norm{f\!\para{t,y,d(t)\eta(x)} - f\!\para{t,y,d(t)\eta(y)}}_X \\
		&\leq L_f^1(C,R)\!\norm{x - y}_X + L_f^2(C,R)\abs{\eta(x) - \eta(y)} \\
		&\leq L_{\overline{f}}(C) \norm{x - y}_X,
	\end{align*}
	where we used Assumption \ref{A1} \ref{A1:first} and \ref{A1:second} in the third step. In the fourth step we  used the Lipschitz continuity of $\eta$.
	The Lipschitz constant $L_{\overline{f}}$ we define in the last step depends on $C$, $R$, and the Lipschitz constant of $\eta$. Note that $R$ and the Lipschitz constant of $\eta$ also depend on $C$. Therefore, $L_{\overline{f}}$ can be expressed in terms of $C$ only.
	
	The proof for $\overline{g}_i$ follows analogously.
\end{pf}

\begin{lemma}\label{lem:Lipschitzcontinuityflow}
	Assume that system \eqref{eq:feedbackImpulsiveSystem} is robustly forward complete. Let $\overline{f}$ be Lipschitz continuous in the second variable on bounded subsets of $X$ with constant $L_{\overline{f}}$ uniformly with respect to the third argument and for all $t \in I$.
	Let for every $i \in\N$, the function $\overline{g}_i$ be Lipschitz continuous in the first variable on bounded subsets of $X$ uniformly with respect to the second argument. Then, on compact time intervals, the solutions of \eqref{eq:feedbackImpulsiveSystem} are Lipschitz continuous with respect to the initial values in bounded subsets of $X$.
\end{lemma}
\begin{pf}
	Let $C > 0$, $\tau > t_0$, $d \in D$ and $x_0, y_0 \in \ball_X(C)$. By Hille-Yosida theorem, there exist constants $M > 0$ and $\lambda \in \R$ such that $\norm{T(t - t_0)}_{\lin(X)}  \leq M e^{\lambda (t-t_0)}$ for all $t \geq t_0$, where $\norm{\ph}_{\lin(X)}$ is the norm for linear operators over $X$.
	
	Let $t \in [t_0, t_1)$ such that $t \leq \tau$. The definition of the mild solution \eqref{eq:ImpulsiveSystemIntegralForm} gives us that
	\begin{align*}
		\overline{x}(t) = T(t - t_0) x_0 + \int_{t_0}^t T(t - s) \overline{f}\!\para{s, \overline{x}(s) ,d(s)} \,\diff s
	\end{align*}
	holds. Let $x_i^- := \overline{x}(t_i^-)$, $x_i := x(t_i) = \overline{g}_i(x_i^-)$ such that
	\begin{align*}
		\overline{x}(t) = T(t - t_i) x_i + \int_{t_i}^t T(t - s) \overline{f}\!\para{s, \overline{x}(s), d(s)} \,\diff s
	\end{align*}
	holds true for all $t \in [t_i, t_{i + 1})$, $t \leq \tau$.
	From this follows that for any two solutions $\overline{x}$ and $\overline{y}$ of \eqref{eq:feedbackImpulsiveSystem} with initial values $x_0$ and $y_0$, respectively, we have
	\begin{align*}
		&\norm{\overline{x}(t) - \overline{y}(t)}_X 
		\leq Me^{\lambda (t - t_i)} \norm{x_i - y_i}_X \\*
		&+ \int_{t_i}^t M e^{\lambda (t - s)} L_{\overline{f}}\!\para{K(C,\tau)} \norm{\overline{x}(s) - \overline{y}(s)}_X \,\diff s,
	\end{align*}
	where $t \in [t_i, t_{i + 1})$, $t \leq \tau$, $y_i^- := \overline{y}(t_i^-)$, $y_i := \overline{y}(t_i) = \overline{g}_i(y_i^-)$, and $L_{\overline{f}}$ is the constant from Lemma \ref{lem:LipschitzcontinuityRHS} in the Appendix and $K = K(C,\tau)$ as introduced in Definition~\ref{def:StabilityDefinitions}~\ref{def:robustForwardCompleteness}. Using Gronwall's inequality, this means
	\begin{align*}
		\norm{\overline{x}(t) - \overline{y}(t)}_X \leq M \norm{x_i - y_i}_X e^{\para{M L_{\overline{f}}\!\para{K(C,\tau)} + \lambda}(t - t_i)}
	\end{align*}
	for all $t \in [t_i,t_{i+1}), t \leq \tau$. The value for $t_{i + 1}$ is bounded by
	\begin{align*}
		&\norm{\overline{x}(t_{i+1}) - \overline{y}(t_{i+1})}_X \\
		&\leq L_{\overline{g}_{i+1}}\!\para{K(C,\tau)} \norm{\overline{x}^-(t_{i + 1}) - \overline{y}^-(t_{i + 1})}_X \\
		&\leq L_{\overline{g}_{i+1}}\!\para{K(C,\tau)}\! M\! \norm{x_i - y_i}_X \!e^{\para{M L_{\overline{f}}\!\para{K(C,\tau)} + \lambda}(t_{i+1} - t_i)},
	\end{align*}
	where $L_{\overline{g}_{i+1}}$ is defined as in Lemma \ref{lem:LipschitzcontinuityRHS} in the Appendix.
	By induction, we obtain
	\begin{align*}
		\norm{\overline{x}(t) - \overline{y}(t)}_X&\leq M \prod\nolimits_{j = 1}^{n}\para{L_{\overline{g}_j}\!\para{K(C,\tau)} M} \norm{x_0 - y_0}_X \\
		&\quad \times e^{\para{M L_{\overline{f}}\!\para{K(C,\tau)} + \lambda}(t - t_0)}
	\end{align*}
	for $t \in [t_n, t_{n+1})$, $ t \leq \tau$ and $n \in \N_0$, where we define the product $\prod_{j = 1}^0 a_j := 1$.
	From this, we can conclude by extreme value theorem that
	\begin{align*}
		&\norm{\overline{x}(t) - \overline{y}(t)}_X \\*
		&\leq \norm{x_0 - y_0}_X \max\nolimits_{t \in [t_0, \tau]} \! \left\{M \prod\nolimits_{j = 1}^{n}\para{L_{\overline{g}_j}\!\para{K(C,\tau)} M}\right. \\*
		&\qquad \left.\vphantom{\prod\nolimits_{j = 1}^{n}}
		\times e^{\para{M L_{\overline{f}}\!\para{K(C,\tau)} + \lambda}(t - t_0)} \, \middle| \,  n \in \N_0: t \in [t_n, t_{n+1})\right\}. 
	\end{align*}
     This gives us a Lipschitz constant for the solutions of \eqref{eq:feedbackImpulsiveSystem} with respect to the initial conditions.
\end{pf}

\begin{lemma}\label{lem:LipschitzContinuityInTime}
	Let system \eqref{eq:feedbackImpulsiveSystem} be robustly forward complete. Let $\overline{f}$ be Lipschitz continuous in the second variable on bounded subsets of $X$ uniformly with respect to the third argument and for all $t \in I$. Then its solutions are locally Lipschitz continuous on intervals $t \in [t_i,t_{i + 1})$, $i \in N_0$, where the Lipschitz constant depends on $\norm{x_0}_X$.
\end{lemma}
The proof follows from \cite[Lemma 4.6]{Mironchenko2018}.

%}

\begin{proposition}\label{prop:Lipschitzcontinuity}
	Let Assumption \ref{A1} hold, and system \eqref{eq:ImpulsiveSystem} be WURS. Then, for any $\eta$ as defined in Definition \ref{def:WURS}, the closed-loop system \eqref{eq:feedbackImpulsiveSystem} has robustly forward complete solutions, which are  Lipschitz continuous with respect to initial values in bounded subsets of $X$ and locally Lipschitz continuous on intervals $t \in [t_i,t_{i + 1})$, $i \in N_0$, where the Lipschitz constant depends on the norm of the initial condition $\norm{x_0}_X$.
\end{proposition}
%\task{Due to space restriction, we do not provide a proof here. For the proof, we refer to ...}

%{
    %\color{green}
    \begin{pf}%{PROOF OF PROP. \ref{prop:Lipschitzcontinuity}}
    	The solutions of system \eqref{eq:ImpulsiveSystem} exist for every $t \geq t_0$. This implies that the solutions of system \eqref{eq:feedbackImpulsiveSystem} also exist for every $t \geq t_0$. Let $\eta$ be the function constructed in Definition \ref{def:WURS}. According to Definition \ref{def:WURS}, system \eqref{eq:feedbackImpulsiveSystem} is UGAS. Therefore, for all $C>0$ and all $\tau > t_0$
    	\begin{align*}
    		\sup\nolimits_{x_0 \in \ball_X(C),\, d \in D,\, t \in [t_0,\tau]} \norm{\overline{x}_d (t;t_0,x_0)}_X \leq \beta(C,0)
    	\end{align*}
    	exists, and system \eqref{eq:feedbackImpulsiveSystem} is robustly forward complete. We apply Lemma \ref{lem:LipschitzcontinuityRHS} from the Appendix  to obtain that $\overline{f}$ and $\overline{g}_i$ are Lipschitz continuous in space on bounded subsets of $X$ and uniformly with respect to $t$ and $d$. Therefore, the preconditions of Lemma \ref{lem:Lipschitzcontinuityflow} in the Appendix are fulfilled, i.e., system \eqref{eq:feedbackImpulsiveSystem} has a solution, which is Lipschitz continuous with respect to the initial values in bounded subsets of $X$. From Lemma \ref{lem:LipschitzContinuityInTime} in the Appendix, it follows that the trajectories are locally Lipschitz continuous on intervals $t \in [t_i,t_{i + 1})$, $i \in N_0$, as claimed in the Proposition.
    \end{pf}
%}

\begin{lemma}\label{lem:WURStoISS}
    Let there exist a UGAS-Lyapunov function $V$ for  system \eqref{eq:feedbackImpulsiveSystem} as given in Definition \ref{def:LyapunovFunctionFeedback}, which is Lipschitz continuous on bounded balls in space and locally Lipschitz continuous on the intervals $[t_i,t_{i + 1})$, $i \in \N_0$.
	Then, there exists an ISS-Lyapunov function of the form given in Defintion \ref{def:LyapunovFunction}, which is Lipschitz continuous on bounded subsets of $X$, and locally Lipschitz continuous on $I \setminus S$ such that it is right continuous and the left limit exists.
\end{lemma}
\begin{pf}
	The UGAS-Lyapunov function $V$as given in Definition \ref{def:LyapunovFunctionFeedback} immediately implies that the inequalities \eqref{ineq:LyapunovFlow}--\eqref{ineq:LyapunovJump} hold for all $x \in X$ and $u \in \ball_U(\eta(x))$. As a consequence of Definition \ref{def:WURS}, \eqref{ineq:LyapunovFlow}--\eqref{ineq:LyapunovJump} hold for all $x \in X$ and $u \in \ball_U\!\para{\psi\!\para{\norm{x}_X}}$. As $\psi$ is invertible, we can set $\chi := \alpha_1 \circ \psi^{-1}$. By this, $V$ satisfies  \eqref{ineq:LyapunovFlow}--\eqref{ineq:LyapunovJump} without additional restrictions.
	
	It remains to show that the third property of Definition~\ref{def:LyapunovFunction} is fulfilled. Let $V\para{t_i,x} < \chi\!\para{\norm{u}_U}$ hold, it follows from \eqref{ineq:LyapunovDefiniteness} that
	\begin{align} \label{ineq:estimateofxinu}
		\norm{x}_X \leq \para{\alpha_1^{-1} \circ \chi}\!\para{\norm{u}_U} = \kappa\!\para{\norm{u}_U},
	\end{align}
	where we define $\kappa \in \mathcal{K}_\infty$, $\kappa(s) = \para{\alpha_1^{-1} \circ \chi}\!(s)$. For $\norm{x}_X \leq C$ and $\norm{u}_U \leq D$, we have
	\begin{align}
		&\norm{g_i(x,u)}_X
		\leq \norm{g_i(x,u)\! - \!g_i(0,u)}_X\! + \!\norm{g_i(0,u)\! - \!g_i(0,0)}_X \nonumber\\
		&\leq L_{g_i}^1(C,D)\norm{x}_X + L_{g_i}^2(C,D)\norm{u}_U \nonumber\\
		&\leq L_{g_i}^1(C,D)\kappa\!\para{\norm{u}_U\!}\!+\! L_{g_i}^2(C,D)\!\norm{u}_U\! =\! \tilde{\alpha}_3\!\para{\norm{u}_U\!}\!, \label{ineq:estimateg_i}
	\end{align}
	where the second inequality applies Assumption \ref{A1} and the last equality follows from the fact that $g_i(0,0) = 0$ by \eqref{eq:feedbackImpulsiveSystem} being UGAS. Note that $D$ is linear in $\norm{u}_U$ and $C$ is linear in $\norm{x}_X$. Therefore, by \eqref{ineq:estimateofxinu}, $L_{g_i}^1$ and $L_{g_i}^2$ are weakly growing functions in $\norm{u}_U$. It can be seen that $\tilde{\alpha}_3(s) := L_{g_i}^1\!\para{\kappa^{-1}(s),s} \kappa(s)+ L_{g_i}^2\!\para{\kappa(s),s} s$ belongs to class $\mathcal{K}_\infty$. By defining the $\mathcal{K}_\infty$-function $\alpha_3 := \alpha_2\circ \tilde{\alpha}_3$ we conclude
	\begin{align*}
		V(t_i,g_i(x,u)) \leq \alpha_2\!\para{\norm{g_i(x,u)}_X} \leq \alpha_3\!\para{\norm{u}_U}.
	\end{align*}
	We used \eqref{ineq:LyapunovDefiniteness} in the first inequality and \eqref{ineq:estimateg_i} in the second one.	So, $V$ is an ISS-Lyapunov function for system \eqref{eq:ImpulsiveSystem}.
\end{pf}

\bibliography{bibliography.bib}

\begin{thebibliography}{12}
\providecommand{\natexlab}[1]{#1}
\providecommand{\url}[1]{\texttt{#1}}
\providecommand{\urlprefix}{URL }
\expandafter\ifx\csname urlstyle\endcsname\relax
  \providecommand{\doi}[1]{doi:\discretionary{}{}{}#1}\else
  \providecommand{\doi}{doi:\discretionary{}{}{}\begingroup
  \urlstyle{rm}\Url}\fi

\bibitem[{Ahmed(2003)}]{Ahmed2003a}
Ahmed, N.U. (2003).
\newblock Existence of optimal controls for a general class of impulsive
  systems on {B}anach spaces.
\newblock \emph{SIAM Journal on Control and Optimization}, 42(2), 669--685.

\bibitem[{Antunes et~al.(2013)Antunes, Hespanha, and Silvestre}]{Antunes2013}
Antunes, D., Hespanha, J.P., and Silvestre, C. (2013).
\newblock Stability of networked control systems with asynchronous renewal
  links: An impulsive systems approach.
\newblock \emph{Automatica}, 49(2), 402--413.

\bibitem[{Bachmann et~al.(2022)Bachmann, Ahmed, and Bajcinca}]{Bachmann2022}
Bachmann, P., Ahmed, S., and Bajcinca, N. (2022).
\newblock Characterization of input-to-state stability for infinite-dimensional
  impulsive systems.
\newblock \emph{{IEEE} Transactions on Automatic Control}.
\newblock Unpublished.

\bibitem[{Dashkovskiy and Mironchenko(2013)}]{Dashkovskiy2013}
Dashkovskiy, S. and Mironchenko, A. (2013).
\newblock Input-to-state stability of nonlinear impulsive systems.
\newblock \emph{{SIAM} Journal on Control and Optimization}, 51(3), 1962--1987.

\bibitem[{Dashkovskiy and Slynko(2021)}]{Dashkovskiy2021a}
Dashkovskiy, S. and Slynko, V. (2021).
\newblock Stability conditions for impulsive dynamical systems.
\newblock \emph{Mathematics of Control, Signals, and Systems}.

\bibitem[{Grüne and Kellett(2014)}]{Gruene2014}
Grüne, L. and Kellett, C.M. (2014).
\newblock {ISS}-{L}yapunov functions for discontinuous discrete-time systems.
\newblock \emph{{IEEE} Transactions on Automatic Control}, 59(11), 3098--3103.

\bibitem[{Guan et~al.(2012)Guan, Wu, and Feng}]{Guan2012}
Guan, Z.H., Wu, Y., and Feng, G. (2012).
\newblock Consensus analysis based on impulsive systems in multiagent networks.
\newblock \emph{{IEEE} Transactions on Circuits and Systems I: Regular Papers},
  59(1), 170--178.

\bibitem[{Hespanha et~al.(2008)Hespanha, Liberzon, and Teel}]{Hespanha2008}
Hespanha, J.P., Liberzon, D., and Teel, A.R. (2008).
\newblock Lyapunov conditions for input-to-state stability of impulsive
  systems.
\newblock \emph{Automatica}, 44(11), 2735--2744.

\bibitem[{Karafyllis and Jiang(2011)}]{Karafyllis2011}
Karafyllis, I. and Jiang, Z.P. (2011).
\newblock \emph{Stability and Stabilization of Nonlinear Systems}.
\newblock Springer-Verlag London.

\bibitem[{Mironchenko and Wirth(2018)}]{Mironchenko2018}
Mironchenko, A. and Wirth, F. (2018).
\newblock Non-coercive {L}yapunov functions for infinite-dimensional systems.
\newblock \emph{Journal of Differential Equations}, 266(11), 7038--7072.

\bibitem[{Rivadeneira and Moog(2015)}]{Rivadeneira2015}
Rivadeneira, P.S. and Moog, C.H. (2015).
\newblock Observability criteria for impulsive control systems with
  applications to biomedical engineering processes.
\newblock \emph{Automatica}, 55, 125--131.

\bibitem[{Sontag(1989)}]{Sontag1989}
Sontag, E.D. (1989).
\newblock Smooth stabilization implies coprime factorization.
\newblock \emph{{IEEE} Transactions on Automatic Control}, 34(4), 435--443.

\end{thebibliography}
\end{document}